\documentclass[runningheads]{llncs}
\usepackage[T1]{fontenc}
\usepackage{graphicx}
\usepackage{amsmath,amssymb,amsfonts}
\usepackage{booktabs}
\usepackage{multirow}
\usepackage{array}
\usepackage{enumitem}
\usepackage{url}
\usepackage{xcolor}
\usepackage[numbers,sort&compress]{natbib}
\usepackage{float}\usepackage{hyperref}
\usepackage{xurl}          
\usepackage{fancyhdr}
\fancypagestyle{plain}{%
  \fancyhf{}
  
  \fancyhead[L]{\footnotesize This is a pre-print version of the paper accepted in ICONIP 2026.}
  \fancyfoot[C]{\thepage}
}

\fancypagestyle{empty}{%
  \fancyhf{}
  
  \fancyhead[L]{\footnotesize This is a pre-print version of the paper accepted in ICONIP 2026.}
  \fancyfoot[C]{\thepage}
}
\hypersetup{hidelinks}     
\begin{document}

\title{GRASP: Graph-Retrieval Automated Scoring Pipeline for Label-Free Multi-Topic Essay Grading}
\titlerunning{GRASP: Graph-Retrieval Automated Scoring Pipeline}
\author{Aafreen Husain\and
Samar Shailendra \and
Saad Sajid Hashmi}
\authorrunning{A. Husain et al.}
\institute{Melbourne Institute of Technology,  Australia}

\maketitle              

\begin{abstract}
Automated short-answer grading research has historically focused on exams consisting solely of questions pertaining to a single topic. Automatic grading of exams containing questions about more than one topic remains less explored. In this work, a Graph-Retrieval Automated Scoring Pipeline (GRASP) is introduced for grading label-free multi-topic science exams. Label-free exams are short-answer exams in which a student's responses to several distinct topics are merged into a single paragraph, with no markup labels or segmentation indicating which span answers which question. Reference answers for each question are encoded into a FAISS vector index via Sentence-BERT, and a semantic similarity graph is constructed over this set of reference answers. At grading time, sentence count heuristics, with a large language model used to resolve ambiguous cases, are first applied to predict how many distinct topics were answered in the student essay. This process is performed without training data or domain-specific example essays. Candidate reference nodes, each storing one (question, reference answer, concatenation of both) from the reference index, are then retrieved through cosine similarity based Retrieval-Augmented Generation (RAG) and Graph Retrieval-Augmented Generation (GRAG). GRAG operates by taking the top cosine matches as seed nodes and then performing a graph traversal over strong edges to find additional reference nodes that may have been missed by RAG. The Hungarian algorithm is then used to optimally assign one reference node per question segment such that no reference is duplicated. Each segment is then graded against its assigned reference independently using GPT-4.1-mini. This experiment is performed to show the effect of retrieval quality on grading accuracy and the benefit of graph-augmented retrieval versus strict cosine similarity methods at various levels of essay complexity.

\end{abstract}
\section{Introduction}
\label{sec:intro}
Consider an automated examination system that generates science exams. Each question it can ask comes from a list of reference nodes it stores. Each node has an exam question as well as the answer to that question. Questions will range from Physics to Chemistry to Biology to Earth Science. To assess students' understanding at the sub-topic level, the system builds a larger question of related $n$ sub-topics randomly selected from its database. For example, it might pull three questions about mineral hardness and then paraphrase them into one exam question. The student is given this question and must answer the $n$ topics in one running paragraph. There are no numbered parts, no section headings, and no markers that can be used to tell where the response to one topic ends and another begins. 

This task is challenging because the $n$ topics were all sampled from the same section of the database and are thus grounded in the same scientific concept. Their reference answers share wording and logic, and the embeddings of the gold reference nodes lie close together in vector space. An untrained retrieval model, when queried for $n$ matching references, may return $n$ copies of the same reference rather than $n$ distinct ones. Retrieval thus faces a three-fold test:
\begin{enumerate}[label=(\roman*)]
\item There is never explicit mention of how many topics $n$ there are. This must be parsed from the student's submission and exam question.
\item Answers can contain multiple sentences, yet there are no annotations connecting a sentence to the topic it addresses. The system must determine which sentence answers which topic based solely on semantics.
\item Sentences must be matched to the appropriate reference node from the pool of reference nodes with no annotated guidance.
\end{enumerate}

Existing automated essay scoring (AES) systems~\cite{xue_hierarchical_2021,amin_enhancing_2025} assume one topic per response and score against a single known reference. They cannot segment student's answer to $n$ topics asked in the question. RAG-based grading systems~\cite{qiu_stella_2025,arslan_survey_2024} retrieve evidence before scoring but pull from a flat index with no awareness of how references relate to each other. This causes retrieval accuracy to collapse when there are multiple related nodes which must all be found simultaneously.

This paper introduces the GRASP pipeline. This pipeline addresses these limitations. 
First, GRASP pipeline detects how many topics question and student answer contains. It splits the answer and the question into topic-focused segments so that each part can be graded against the reference it actually addresses. Secondly, it organises reference nodes into a graph that captures how they relate to one another. This allows related references to be retrieved together rather than from a flat, unaware index. Finally, it assigns each segment to its most appropriate reference node through an optimal global matching step. This ensures that distinct but semantically similar references are kept apart rather than collapsed into near-duplicates. Our main contribution is this end-to-end problem formulation for label-free multi-topic grading, together with the
evaluation methodology that separates retrieval quality from grading quality. The graph expansion mechanism itself adapts existing GraphRAG-style retrieval \cite{edge_local_2025, hu_grag_2025} to this new
setting rather than introducing a new retrieval algorithm.

The remainder of this paper is structured as follows. Section~\ref{sec:literature} reviews related work. Section~\ref{sec:problem} formally defines the multi-topic grading problem. Section~\ref{sec:methodology} describes the GRASP pipeline. Section~\ref{sec:results} presents results and analysis. Section~\ref{sec:conclusion} concludes with directions for future work.

\section{Literature Survey}
\label{sec:literature}

Research on automated short-answer and essay grading spans five interconnected thematic areas. This survey covers foundational and recent work across AES models, semantic similarity, RAG-based systems, graph-enhanced retrieval, and LLM-based feedback generation.

\subsection{Automated Essay Scoring (AES) Models}

Transfer learning using pre-trained transformer models has dominated AES research over the past decade for holistic scoring and multi-dimensional scoring tasks. Xue et al.~\cite{xue_hierarchical_2021} introduce hierarchical BERT-based transfer learning approach for multidimensional essay scoring which attained improvements of 4.5\% Quadratic Weighted Kappa (QWK) on ASAP dataset and 8.1\% on CELA dataset by segmenting essays into hierarchical representations and attending over segments using attention pooling. Amin et al.~\cite{amin_enhancing_2025} show few-shot transformer based AES can attain QWK=0.97 and QWK=0.94 for holistic scoring and content scoring respectively on ASAP with little supervision, but the model was not interpretable and raised concerns regarding scoring bias. Wang et al.~\cite{wang_use_2022} introduce multi-scale BERT which jointly learns document-level, token-level, and segment-level representations via joint training to improve holistic scoring accuracy on ASAP. Sun and Wang~\cite{sun_automatic_2024} use regression heads trained on top of fine-tuned BERT models for scoring across multiple traits (e.g., content, organisation, style) and show competitive results across different scoring dimensions. Do et al.~\cite{do_autoregressive_2024} introduce ArTS, an AES method that frames multi-trait AES as an autoregressive sequence generation problem via T5. By generating scores one trait at a time, they achieve state-of-the-art results on multiple benchmarks. Cummins et al.~\cite{cummins_constrained_2016} introduce constrained multi-task learning to AES, jointly optimising for several scoring objectives under constraint and showing improved generalisation to unseen prompts. Rodriguez et al.~\cite{rodriguez_language_2019} explore the utility of pre-trained language models for AES and conclude that contextual embeddings significantly outperform feature engineered baselines, Elmassry et al.~\cite{elmassry_systematic_2025} provide a systematic review finding this trend to hold true across publications from 2019 to 2023.

\subsection{Semantic Similarity Approaches}

Semantic similarity is the representational paradigm employed by most state-of-the-art short-answer grading systems. Reimers and Gurevych~\cite{reimers_sentence-bert_2019} created Sentence-BERT (SBERT), a framework which fine-tunes BERT via a siamese network architecture to obtain dense fixed-length sentence embeddings that allow for fast cosine similarity searches. The embedding model used for indexing and retrieval in this work is SBERT. Patil and Agrawal~\cite{patil_auto_nodate} used siamese Bi-LSTM networks with attention mechanisms applied to short-answer grading. Their work shows that siamese semantic similarity scoring can correctly pair reference answers with student submissions when applied to narrowly defined domains. Gao et al.~\cite{gao_simcse_2022} introduce SimCSE, a simple contrastive learning framework for sentence embeddings that trains representation models with both supervised and unsupervised contrastive learning objectives. Khattab and Zaharia~\cite{khattab_colbert_2020} introduce ColBERT, an efficient passage retrieval framework built upon contextualized late interaction that allows dense retrieval to scale to large corpora. A disadvantage of calculating sentence similarity via a scalar function such as SBERT is that we are given no causal inference to understand how two passages may be related, graph-augmented approaches seek to solve this shortcoming. Li et al.~\cite{li_enhanced_2022} propose a hybrid neural architecture for AES which extracts linguistic, semantic, and structural attributes of an essay document which are fused over a BERT encoder. Conversely, Janda et al.~\cite{janda_syntactic_2019} showed syntactic, semantic, and sentiment features contributed jointly to overall automated essay score. Vaswani et al.~\cite{vaswani_attention_nodate} introduced the self-attention mechanism.

\subsection{RAG-Based Grading Systems}

Retrieval-augmented generation provides an elegant solution to grounding LLM outputs to factual evidence. Qiu et al.~\cite{qiu_stella_2025} introduce SteLLA, a structured grading framework that leverages RAG with LLMs to accomplish reference-grounded short-answer grading. SteLLA transforms reference grading criteria into question-answer pairs and grounds LLM scoring in retrieved evidence, reaching Cohen's $\kappa = 0.672$ on undergrad biology exams. Arslan et al.~\cite{arslan_survey_2024} review RAG systems utilizing LLMs and empirically show that predicting with relevant evidence passages retrieved beforehand drastically improves factual grounding and decreases hallucination in generated feedback. However, flat retrieval, the method implemented by base RAG models, introduces noisy duplicate passages, lacks the ability to model relationships between reference passages, and doesn't scale to the multi-reference problem of retrieving all $n$ nodes. Dai and Le~\cite{dai_semi-supervised_2015} and Radford et al.~\cite{radford_language_nodate} laid the pre-training groundwork that retrieval-augmented systems utilize for language model knowledge.

\subsection{Graph-Enhanced Retrieval (GRAG)}

Graph-augmented retrieval directly addresses the limitations of flat RAG by organising retrieved knowledge into structured entity-relation graphs. Edge et al.~\cite{edge_local_2025} proposed GraphRAG, which constructs entity-relation graphs from retrieved text to enable multi-hop reasoning over both local subgraphs and global community summaries. This approach is particularly effective for queries requiring synthesis across multiple related documents, such as identifying all reference nodes relevant to a multi-topic essay. Hu et al.~\cite{hu_grag_2025} introduced GRAG: Graph Retrieval-Augmented Generation, which organises retrieved evidence into entity-relation structures to enable structured multi-hop reasoning and reduce retrieval noise from irrelevant passages. Both methods are directly instantiated in the GRASP pipeline: reference nodes are connected by cosine similarity edges, and GRAG retrieval expands from FAISS~\cite{douze_faiss_2025} seed results along strong edges ($\geq 0.70$ weight) to recover reference nodes that are semantically adjacent but not the top-scoring singleton retrievals.

\subsection{LLM-Based Feedback Generation}
Recent research has investigated applications of LLMs towards generating instructionally valuable feedback along with scores. To better adapt LLMs to automatic scoring tasks, Latif and Zhai~\cite{latif_fine-tuning_2023} fine-tuned ChatGPT with reference-aligned annotations and showed that LLMs could be steered towards conforming to marking schemes through instruction tuning. Stahl et al.~\cite{stahl_exploring_2024} examined chain-of-thought prompting and trait-aware prompting techniques for the joint task of scoring and feedback generation, showing that chained prompts significantly improve trait-level scoring coherence. Hussein et al.~\cite{hussein_trait-based_2020} introduced a trait-based deep learning AES system that generates adaptive feedback conditioned on individual student performance profiles to vary the specificity of feedback. Hou et al.~\cite{hou_improve_2025} showed that supplementing LLM prompts with engineered linguistic features narrows performance gaps between small- and large-scale models, indicating that feature-aware prompting may serve as a more economical alternative to fine-tuning. Xiao et al.~\cite{xiao_human-ai_2025} designed a human-AI collaborative scoring framework for essays informed by a dual-process theory, which modularizes rapid holistic scoring from deliberate reference-grounded review. In a cross-disciplinary survey of auto-grading methods with LLMs, Eneye et al.~\cite{frederick_eneye_advances_2025} found fairness, interpretability, and domain adaptation to remain open issues, which aligns with our design of GRASP to be training-data free.

\section{Problem Definition}
\label{sec:problem}

Considering the automated examination system which was introduced in Section~\ref{sec:intro}, the grading problem is described as following constituent tasks:

A \textit{multi-topic essay} is a continuous student text $E$ which is written in response to a question which can include $n$ related topics, where $n \in \{1, 2, 3, 4\}$. The text contains no structural markers such as ``Part 1:'', numbered sections, or line breaks between answers. The answers appear in an arbitrary order that may not correspond to the order of topics on the exam question. Critically, for $n \geq 2$ the topics addressed in any single essay are not chosen at random from the full reference pool. They are required to form a connected subgraph in the reference node similarity graph (see Section~\ref{sec:methodology}). This means that their reference answers are semantically related with pairwise cosine similarity $\geq 0.42$. This constraint reflects realistic exam design. Topics in the same question tend to share a domain or theme. This makes the retrieval problem significantly harder, because the correct reference nodes for a given essay are closer together in embedding space than arbitrary pairs.

A \textit{reference node} is a triple $R_i = (q_i, a_i, t_i)$, where $q_i$ is the original exam question, $a_i$ is the reference correct answer, and $t_i = q_i \oplus a_i$ is the concatenation of both. The full reference index contains $|R| = 79$ nodes drawn from the SciEntsBank dataset, one node per unique (question, reference answer) pair. This includes multiple science topics including physics, chemistry, biology, and earth science.

The \textit{topic detection task} requires the system to predict $n$ from the student essay $E$ together with the paraphrased exam question, with no access to Part~N: labels or any other structural information. The system must infer $n$ from linguistic content alone.

The \textit{segmentation task} requires partitioning $E$ into $n$ non-overlapping, contiguous segments $\{s_1, \ldots, s_n\}$ such that each segment $s_i$ contains the student's complete answer to exactly one topic in the question. Segments must cover the full text with no gaps.

The \textit{retrieval task} operates on a paraphrased version of the exam question as a query and retrieves candidate reference nodes from index $\mathcal{I}$, where $\mathcal{I}$ is built using SBERT~\cite{reimers_sentence-bert_2019} embeddings of reference answers only. This is a cross-modal retrieval design: the query is a paraphrased exam question while the index contains reference answers. Both retrieval configurations return the same number of candidates $k = n + 2$, so that any difference in downstream grading is attributable to retrieval quality rather than to differing candidate-set sizes. Retrieval must identify all $n$ correct reference nodes within these $k$ candidates.

The \textit{assignment task} maps the $n$ paraphrased question segments to the retrieved candidate reference nodes through a one-to-one matching, so that each segment is paired with a distinct reference node and no candidate is reused. This is solved with the Hungarian algorithm. Full details are given in Section~\ref{sec:methodology}.

The \textit{grading task} assigns a score $g_i \in \{1, 2\}$ to each (student answer segment, reference node) pair, where score 2 denotes a correct answer and score 1 denotes a partially correct or incomplete answer. The final essay score is:
\begin{equation}
G(E) = \sum_{i=1}^{n} g_i, \quad G(E) \in [n, 2n]
\label{eq:finalscore}
\end{equation}

To isolate retrieval quality from grading quality, we additionally define a \textit{TRUE Oracle} upper bound that bypasses detection, segmentation, and retrieval entirely: it grades the $n$ gold reference answers directly against their gold reference nodes, achieving full retrieval (N-Hit $=1$) by construction. The Oracle measures the ceiling imposed by the LLM grader alone, against which RAG and GRAG are compared.

Looking at the examination system: for the student who has written three sentences about mineral hardness, the system must identify $n=3$, split the three sentences into three segments, retrieve $k = n+2 = 5$ candidate reference nodes, match each of the three segments to a distinct reference node, and award a score of 1 or 2 per segment. The final score ranges from 3 to 6. The challenge is that all three correct reference nodes were drawn from the same topical cluster in the database , they relate to scratch testing and hardness , so graph-augmented retrieval is necessary to reliably recover all three simultaneously.

\section{Methodology}
\label{sec:methodology}

\begin{figure}[ht]
    \centering
    \includegraphics[width=\textwidth,keepaspectratio]{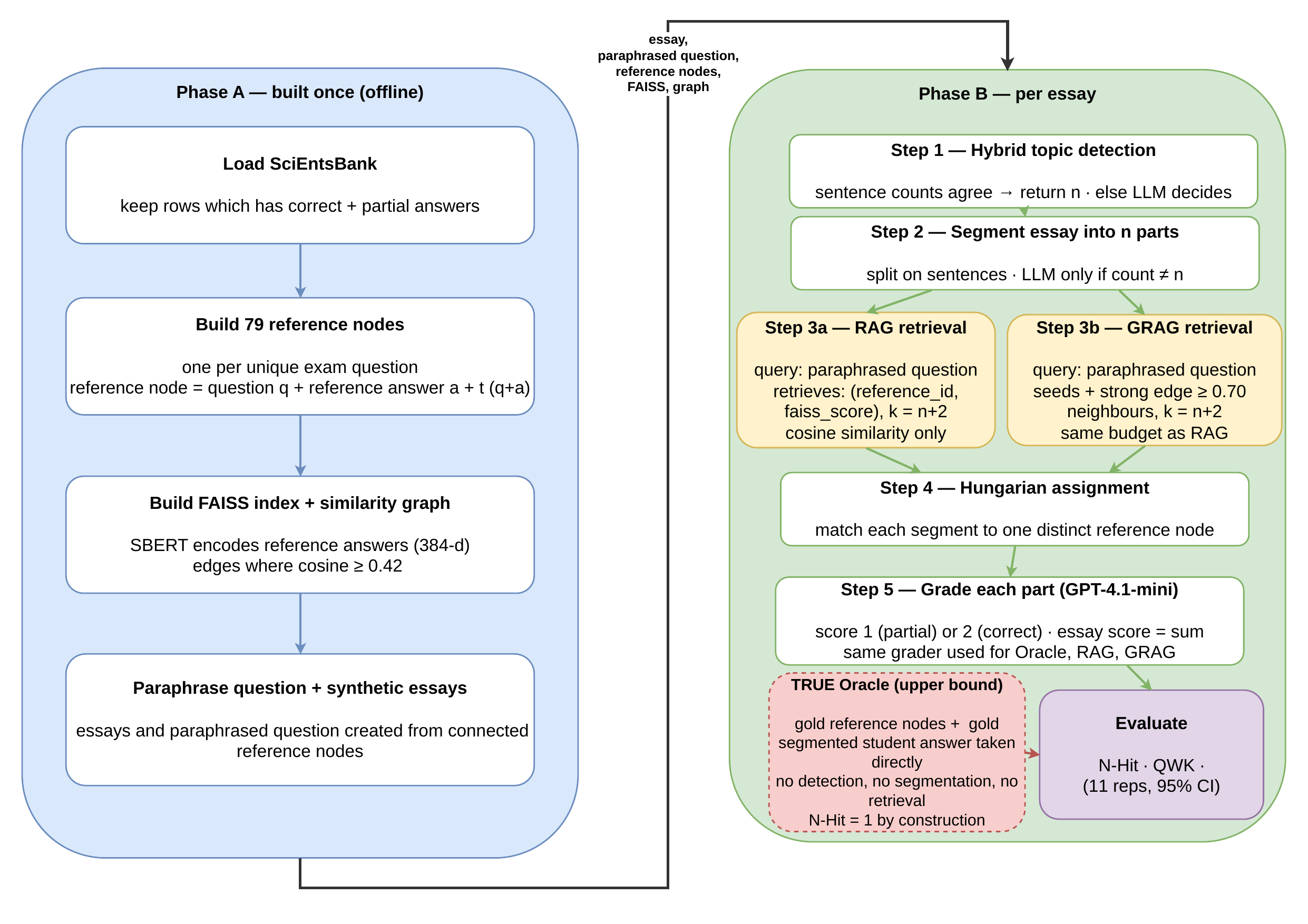}
    \caption{GRASP Pipeline}
    \label{fig:placeholder}
\end{figure}

GRASP grades a label-free multi-topic essay in two phases, as shown in
Fig.~\ref{fig:placeholder}. Phase A covers the offline steps that are run once to
build the dataset, the FAISS index, and the similarity graph. Phase B covers the
per-essay steps run at grading time: detection, segmentation, retrieval,
assignment, and grading. Splitting the work this way lets the expensive corpus
setup be built once and reused, so each essay only triggers the lighter
per-essay steps.

\subsection{Phase A: Offline steps for building the dataset}

This phase generates the static resources against which each essay will be
scored. It runs once, separately from any student submission, and outputs the
reference nodes, FAISS index, and similarity graph that will be used in Phase B.

\textbf{Dataset.} We use the SciEntsBank 
dataset\,\footnote{\url{https://huggingface.co/datasets/nkazi/SciEntsBank}} from HuggingFace~\cite{dzikovska_semeval_2013}. Labels other than \textit{correct} (scored 2) and \textit{partially\_correct\_incomplete} (scored 1) are discarded. Rows for which the diagram filter detected a visual stimulus are filtered out. Rows are dropped where the text component is empty.

\textbf{Build reference nodes.} Reference nodes $R_i = (q_i, a_i, t_i)$ are built, one node per remaining unique (question, reference answer) pair in dataset. Nodes store $Q$, $A$ and concatenated QA pairs $t_i$.

\textbf{Build FAISS index.}  Every reference answer $a_i$ is encoded with SBERT (default \texttt{all-MiniLM-L6-v2}). Answer embeddings are stored in a FAISS inner-product index. Note: index only stores reference answer-form text. Query inputs are always exam questions. Retrieval must bridge the question-answer semantic gap using frozen SBERT embeddings alone. The essay-construction heuristic compounds the difficulty: by construction, all $n$ reference nodes connected in $G$ for a given essay (every essay is guaranteed to have one) have reference answers which are semantically similar. Their embeddings will tend to cluster together in space. This makes it more difficult for naive flat top-$k$ retrieval to recover all $n$ without graph expansion.

\textbf{Build node similarity graph $G = (V, E)$.} Nodes $V$ are simply the set of 79 reference nodes constructed previously. Add an edge $(R_i, R_j) \in E$ weighted $w_{ij}$ iff $\cos(e_{a_i}, e_{a_j}) \geq 0.42$. Edges with weight $w_{ij} \geq 0.70$ are deemed strong and are used later during graph expansion.

\textbf{Build paraphrase cache.} Each reference question $q_i$ is paraphrased once using GPT-4.1-mini, output is stored in cache keyed on reference node, reusing for every essay. This amounts to one LLM call per unique reference node.

\textbf{Construct similarity-constrained topic groups.} Before constructing essays, groups of $n$ reference nodes are preselected from $G$. These groups exhibit strong semantic similarity between every reference node, the topics within each essay should relate to each other, as would be in a real realistic exam domain where the questions on the same test paper are generally drawn from the same domain or theme. For $n = 1$ essays,  each node in its own is a single-topic group. For groups of size $n \geq 2$, nodes must be strongly connected in $G$ , formally speaking, each group forms a clique. Cliques are chosen as follows. For $n = 2$, every edge $(R_i, R_j)$ with weight $\geq 0.42$ is valid so long as questions are distinct. For $n \geq 3$, recursively enumerate all cliques of size $\geq n$ in $G$, then take the size-$n$ subsets. Each group is labelled by its minimum edge weight $w_{\min}$. Groups with high $w_{\min}$ , that correspond to sets of topics that are more semantically coherent  are preferred.

\textbf{Synthetic essay assembly.} For each $n \in \{1, 2, 3, 4\}$ 160 label-free synthetic essays are generated.  Essays are assembled as follows. Let $H$ be the set of reference nodes forming an $n$-topic group. Sample student answers $\forall R_i \in H$ from pool of available sentences for $R_i$, then they are concatenated in node order. This results in  student essay $E$ with no structural tags, separators, or line breaks. Parallel paraphrased question essay is constructed from per-topic paraphrase question from cache. Numeric gold score for essay is computed as sum of gold scores of source $n$ answers.

\subsection{Phase B: Per-Essay Processing Steps}

This phase runs once per student essay. Using the resources built in Phase A, it
applies five steps in sequence: topic detection, segmentation, retrieval,
assignment, and grading.

\textbf{Hybrid topic detection.} Let $s_1$ and $s_2$ respectively denote the sentence count of $E$ and of its paraphrased question. If $s_1$ = $s_2$ , the agreed value is returned immediately as $n$ without calling the LLM. Otherwise, the LLM is called once at temperature 0, with both texts provided as context along with the sentence counts as explicit hints. Prompt is carefully designed to be robust to common failure mode of over-detection, where student writes elaborates the same topic for several consecutive sentences. The hybrid approach skips the LLM call entirely for the majority of essays, where the student sentence count and question sentence count already agree, incurring only a single call on the remaining essays where they disagree.

\textbf{Sentence-first segmentation heuristic.}  $E$ is segmented  by attempting to split on sentence boundaries first. LLM is called if the  resulting segment count $\neq$ detected $n$.  Answer segments and, in parallel, question segments obtained. by splitting the per-topic paraphrases and $E$.

\textbf{Equal-$k$ retrieval.}  Both retrieval methods are configured to have budget $k = n_{\text{detected}} + 2$ computed from detected value of $n$. Extra two candidates gives room for retrieval candidates to ``move around'' during assignment step. Holding budget equal between RAG and GRAG configurations isolates any grading differences and bias.

\textbf{ Retrieval by RAG.} Paraphrased question essay is encoded with SBERT into fixed-dimensional query vector. Top $k = n + 2$ reference nodes are obtained cosine distance from FAISS index.

\textbf{Retrieval by GRAG.} Let seed set $S$ contain the top 3 reference nodes returned by FAISS under exact same settings as RAG. Candidate pool is selected by performing graph expansion out of seeds, following outgoing strong edges ($w \geq 0.70$, one hop) from each seed node to retrieve candidate pool. Each candidate $R_j$ is scored with combined score of $\text{score}(R_j) = \alpha \cdot \cos(e_Q, e_{a_j}) + \beta \cdot \max_{s \in S} w_{sj}$ where $\alpha = 0.80$ and $\beta = 0.20$, where second term corresponds to max over seeds $s$ of strong edge weight between $R_j$ and $s$. As opposed to RAG above, candidates are retrieved using GRAG.

\textbf{Assignment step.} With candidate pool of reference  input,each paraphrased question segment  is embedded independently with SBERT and a cost matrix is constructed. Hungarian algorithm is used to map each paraphrased question segment to a reference node. Both RAG and GRAG assignments use this procedure.

\textbf{TRUE Oracle choice.} 
Each essay is also graded via a TRUE Oracle path that grades the gold reference nodes and answer segments directly, bypassing detection, segmentation, retrieval, and assignment. This gives N-Hit $= 1$ by construction and serves as the upper bound against which RAG and GRAG are compared.

\textbf{LLM grading with score variance.} Each answer segment, reference node pair from assignment procedure is passed to GPT-4.1-mini. Model independently grades segment as 2 (correct) or 1 (partially correct), totaling afterwards for essay score. Grading procedure is repeated 11 times over at temperature 0.7 for Oracle, RAG, and GRAG, while holding the deterministic retrieval and assignment stages constant. Repeating the stochastic grading enables empirical computation of score variance and 95\% confidence intervals for each essay, in addition to the point estimate of mean score.

\section{Results}
\label{sec:results}

Oracle, RAG, and GRAG are evaluated on essays with topic counts $n \in \{1, 2, 3, 4\}$ from SciEntsBank. For each essay, the paragraph question is used as the query to retrieve candidate reference nodes from the reference-answer index. The retrieved candidates are then assigned to the paraphrased question segments via the Hungarian algorithm, and each (student-answer segment, assigned reference, question) triple is graded by the LLM as 2 (correct) or 1 (partially correct). Each essay is graded 11 times at temperature 0.7. The summed essay score ranges from $n$ to $2n$, so the full grading scale spans 1 to 8. We report two metrics: N-Hit (fraction of essays for which retrieval recovered all gold reference nodes) and Quadratic Weighted Kappa (QWK). The Oracle is given the gold reference nodes directly (N-Hit $= 1$ by construction), which isolates grader ability from retrieval. Hybrid topic detection recovered the correct $n$ for 90.8\% of essays.

\subsection{Evaluation Metrics}
We assess the pipeline using two metrics, defined below: the N-Hit rate for retrieval quality and Quadratic Weighted Kappa for grading agreement.

\textbf{N-Hit rate.} For each essay with $n$ topics, the system assigns one
reference node per topic segment. N-Hit is $1$ if all $n$ gold reference nodes
are retrieved in the assigned set and $0$ otherwise. The N-Hit rate is the mean over all essays.

\textbf{Quadratic Weighted Kappa (QWK).} QWK is a statistical measure of the
level of agreement between two raters who assign items to an ordinal
scale~\cite{doewes_evaluating_2023}. It is well suited to ordinal classification
problems, such as integer student scores, and penalises larger disagreements
more heavily than smaller ones. The coefficient ranges from $-1$ to $1$, where
$1$ indicates perfect agreement, with $0$ indicating no agreement better than a random process.

\begin{figure}[!ht]
\centering
\includegraphics[scale=0.35]{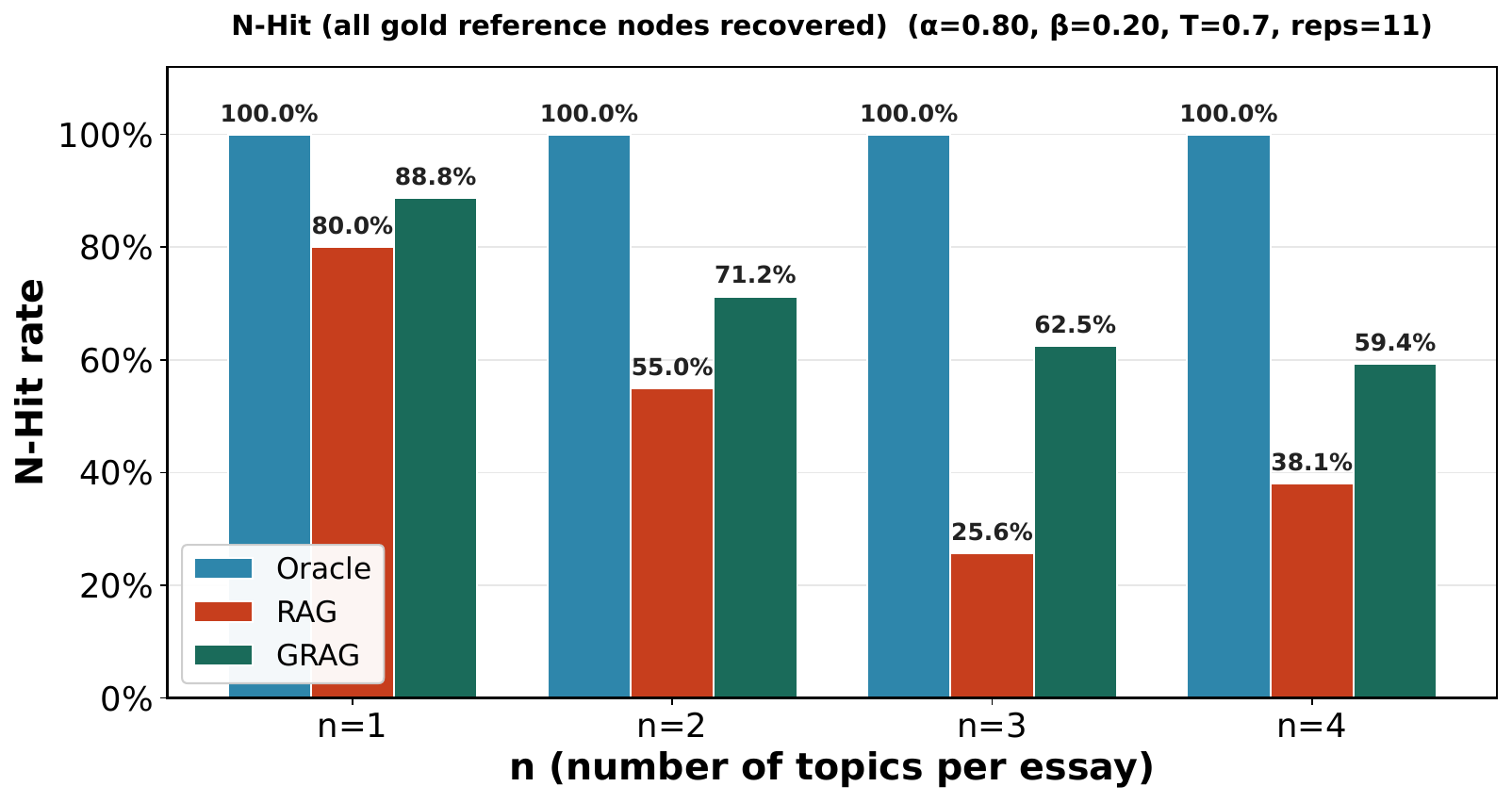}
\caption{N-Hit by topic count $n$: fraction of essays for which retrieval recovered all gold reference nodes. Oracle is perfect by construction. }
\label{fig:nhit}
\end{figure}

\subsection{Retrieval.} GRAG retrieves significantly more gold reference nodes than flat RAG (Fig.~\ref{fig:nhit}). At $n = 1$, 88.8\% vs 80.0\%, at $n = 2$, 71.2\% vs 55.0\%, at $n = 4$, 59.4\% vs 38.1\%. The biggest difference is at $n = 3$, where GRAG attains 62.5\% while RAG drops to 25.6\%. GRAG recovers the complete gold set over twice as frequently. Since an essay's topics arise from a connected subgraph of the topic graph, gold reference nodes are tightly clustered in embedding space. Flat top-$k$ retrieval returns near-duplicates of the seed node, GRAG traces strong graph edges out to the cluster's neighbours. These results imply graph structure benefits the harder examples more.

\subsection{Grading and false positives.} 
\begin{figure}[!ht]
\centering
\includegraphics[scale=0.4]{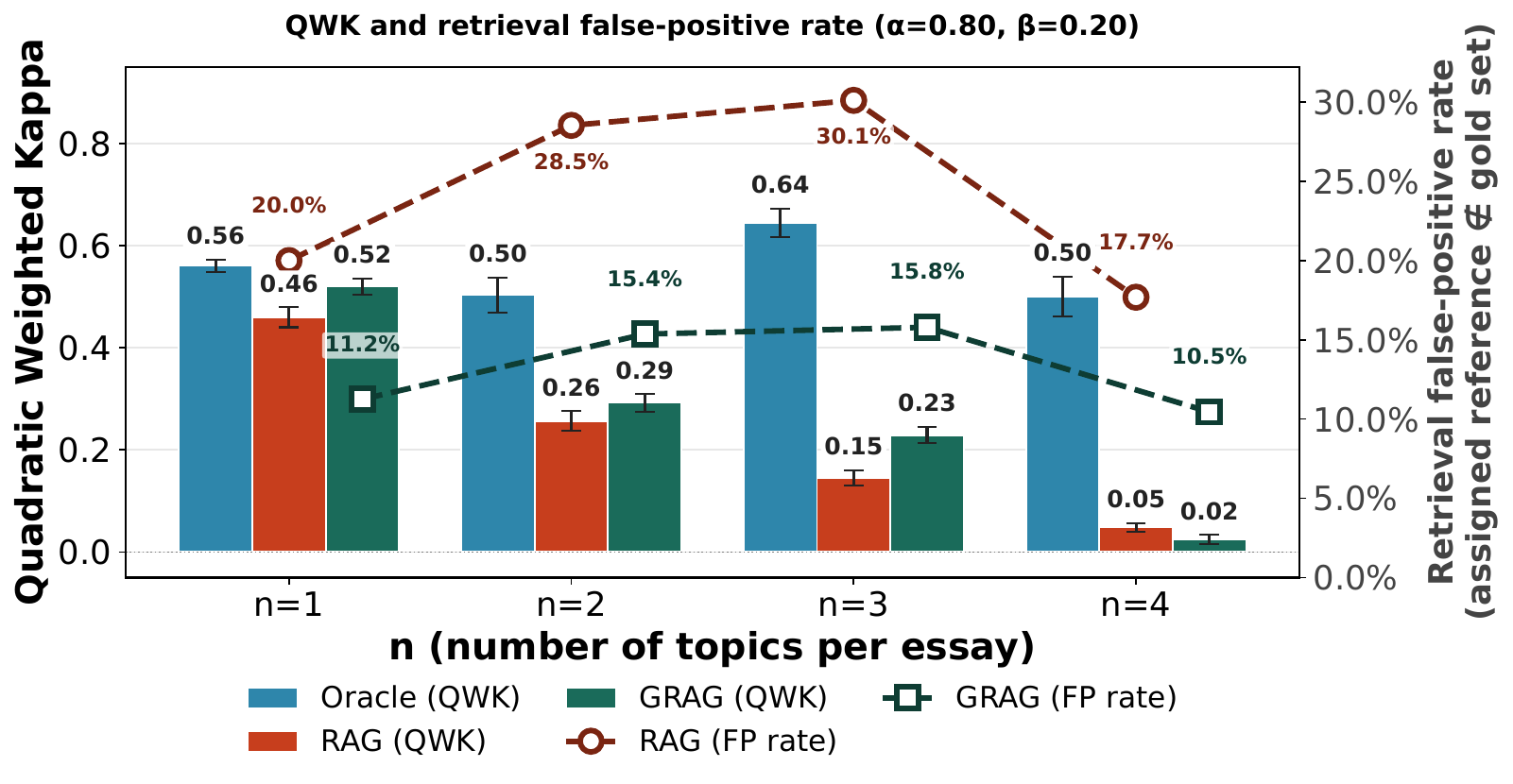}
\caption{Left axis: QWK by topic count $n$ for Oracle, RAG, and GRAG. Right axis: retrieval false-positive rate (assigned reference $\notin$ gold set) for RAG and GRAG.}
\label{fig:qwk_fp}
\end{figure}

A false positive (FP) is a retrieved reference node that is not in the gold set: the grader is shown the wrong reference and asked to mark the student's answer segment against it. The damage is proportional to the semantic closeness of the wrong reference to the correct one. The cleanest case arises from the pitch cluster. The gold reference node asks what happens to pitch when the string is pulled \emph{tighter}, RAG retrieves the reference node about a \emph{shorter} string instead. Both reference answers state only the effect, ``the pitch will be higher'', because the cause (tightness vs shortness) is encoded in the question stem, not in the reference answer. Most students copy only the effect, e.g.\ ``the pitch would get higher'', so the grader, comparing the student's effect-claim against the reference's effect-claim, awards full credit 2 even on the wrong reference. The ambiguity is built into the dataset: four pitch reference nodes share a common effect and vary only on cause.  The FP rate quantifies this exposure (Fig.~\ref{fig:qwk_fp}): RAG runs 17.7--30.1\% across $n = 1$--4, GRAG 10.5--15.8\%. QWK reflects where the wrong references actually hurt: GRAG beats RAG at $n = 1$ (0.52 vs 0.46), $n = 2$ (0.29 vs 0.26), and $n = 3$ (0.23 vs 0.15). At $n = 4$ both RAG and GRAG collapse (Oracle 0.50, GRAG 0.02, RAG 0.05). The grader's own ceiling falls here. Oracle drops from 0.64 at $n = 3$ to 0.50, but the large gap between Oracle and GRAG shows that retrieval and assignment errors pull performance well below even that lowered ceiling. 
We view $n = 4$ grading as an open problem rather than a solved
case: with only four topics compounding retrieval, assignment, and grading error at once, this setting is the clearest evidence that GRASP, and graph-augmented retrieval more broadly, has not yet closed the gap to reliable grading at the highest topic counts we tested.
All QWK values are reported as the mean over 11 grading repetitions, with 95\% confidence intervals shown as error bars in Fig.~\ref{fig:qwk_fp}.

\subsection{Why we report per-$n$.}
Pooled across all $n$, QWK is essentially flat (Oracle 0.978, RAG 0.892,
GRAG 0.901 at $\alpha = 0.8$, $\beta = 0.2$). This is an instance of Simpson's
paradox: $n$ is correlated with both predictor and outcome, since higher-$n$
essays carry higher gold totals, and any grader that approximately tracks $n$
inherits that correlation, inflating pooled agreement on the 1--8 scale. We
report both pooled and per-$n$ results because each answers a different
question, and both are valid. At test time the system is not told $n$, it must
detect it from the essay and question alone (recovered correctly for 90.8\% of
essays). The pooled figure therefore reflects genuine end-to-end performance,
where correctly inferring how many parts an essay has is itself part of the task
and legitimately contributes to agreement. The per-$n$ figures hold $n$ fixed
and so isolate the retrieval contribution, exposing the differences between
methods that the pooled number masks. Neither is an artifact to be discarded: the pooled result measures the full deployed pipeline, while the per-$n$ result
attributes credit to retrieval, and these are reported together throughout.
These results imply that GRAG's higher per-$n$ QWK reflects a genuine retrieval
improvement rather than an effect of the wider score range, since holding $n$
fixed removes the score-range advantage the pooled figure carries.

\subsection{Limitations}
Our evaluation essays are synthetic: they are assembled by concatenating sentences from separately authored, single-topic SciEntsBank reference answers, in the fixed order of their source reference nodes, and the paraphrased exam questions are generated from the same underlying reference questions. This construction guarantees clean ground truth for topic count, segmentation boundaries, and gold reference nodes, which real multi-topic student essays do not provide, but it does not capture ways genuine students blend topics within a single sentence, use cross-referential language between topics, or answer topics out of the order in which they were asked. As a result, the detection, segmentation, and retrieval figures reported here should be read as an upper bound relative to real deployment, and validating the pipeline on genuine multi-topic student responses, including answers presented out of order, is an important direction for future work.

\section{Conclusion and Future Work}
\label{sec:conclusion}

 We presented GRASP, a label-free pipeline for grading synthetically constructed multi-topic science essays designed to approximate realistic multi-topic exam settings. In it, a graph-augmented retrieval module called GRAG recovers reference nodes left behind by flat similarity retrieval. Between one and four topics per essay, GRAG retrieved the entire gold set of references significantly more frequently than flat RAG at every $n$, and by wider margins as the gold set became more tightly clustered. The effect was starkest at $n=3$, where GRAG retrieved the full gold set more than twice as often as its RAG baseline. This translated to higher grading agreement at low and moderate topic counts. 
 By supplementing our experiments with a TRUE Oracle that retrieved all references perfectly, we isolated grading errors at higher topic counts from retrieval errors: even under perfect retrieval, QWK falls to 0.50 at n = 4, indicating that grading quality and retrieval quality are at least partially separable. This pattern is consistent with a ceiling imposed by the LLM grader itself, though we tested only GPT-4.1-mini as the grader and cannot rule out that a stronger or differently-prompted model would narrow this gap; we leave a multi-grader comparison to future work. Taken together, our experiments point to a narrow but reliable conclusion: given a set of gold references clustered together semantically, access to graph structure recovers them more reliably than does embedding similarity.

GRAG's largest gains came where flat embedding retrieval failed most noticeably. When the relevant reference nodes sat close together in embedding space and gold-set similarity scores alone were insufficient to pick out the correct references from their near-duplicates. This suggests a larger application for our method. While we used document similarity to build a graph connecting reference nodes, that graph was only one way of connecting those nodes, and our retrieval mechanism can operate over any connections an examiner chooses to draw. The best connections aren't necessarily those implied by SBERT similarity scores. Consider a co-occurrence graph that connects reference nodes that appear together on real exams/course curriculum rather than ones that simply have similar reference answers. Exploring such alternative graph constructions is a promising direction
for future work.

\setlength{\bibhang}{1em}
\bibliographystyle{splncs04}
\bibliography{references}

@article{xue_hierarchical_2021,
  title = {A {Hierarchical} {BERT}-{Based} {Transfer} {Learning} {Approach} for {Multi}-{Dimensional} {Essay} {Scoring}},
  author = {Xue, Jin and Tang, Xiaoyi and Zheng, Liyan},
  journal = {IEEE Access},
  volume = {9},
  pages = {125403--125415},
  year = {2021},
  doidoi = {10.1109/ACCESS.2021.3110683},
  urlurl = {https://ieeexplore.ieee.org/abstract/document/9530411},
  issn = {2169-3536},
}

@article{amin_enhancing_2025,
  title = {Enhancing {Essay} {Scoring}: {An} {Analytical} and {Holistic} {Approach} {With} {Few}-{Shot} {Transformer}-{Based} {Models}},
  author = {Amin, Tahira and Tanoli, Zahoor-Ur-Rehman and Aadil, Farhan and Awan, Khalid Mahmood and Lim, Sangsoon},
  journal = {IEEE Access},
  volume = {13},
  pages = {12483--12501},
  year = {2025},
  doidoi = {10.1109/ACCESS.2025.3530272},
  urlurl = {https://ieeexplore.ieee.org/document/10843186/},
  issn = {2169-3536},
}

@misc{edge_local_2025,
  title = {From {Local} to {Global}: {A} {Graph} {RAG} {Approach} to {Query}-{Focused} {Summarization}},
  author = {Edge, Darren and Trinh, Ha and Cheng, Newman and Bradley, Joshua and Chao, Alex and Mody, Apurva and Truitt, Steven and Metropolitansky, Dasha and Ness, Robert Osazuwa and Larson, Jonathan},
  year = {2025},
  publisher = {arXiv},
  doidoi = {10.48550/arXiv.2404.16130},
  urlurl = {http://arxiv.org/abs/2404.16130},
}

@misc{hu_grag_2025,
  title = {{GRAG}: {Graph} {Retrieval}-{Augmented} {Generation}},
  author = {Hu, Yuntong and Lei, Zhihan and Zhang, Zheng and Pan, Bo and Ling, Chen and Zhao, Liang},
  year = {2025},
  publisher = {arXiv},
  doidoi = {10.48550/arXiv.2405.16506},
  urlurl = {http://arxiv.org/abs/2405.16506},
}

@misc{wang_use_2022,
  title = {On the {Use} of {BERT} for {Automated} {Essay} {Scoring}: {Joint} {Learning} of {Multi}-{Scale} {Essay} {Representation}},
  author = {Wang, Yongjie and Wang, Chuan and Li, Ruobing and Lin, Hui},
  year = {2022},
  publisher = {arXiv},
  doidoi = {10.48550/arXiv.2205.03835},
  urlurl = {http://arxiv.org/abs/2205.03835},
}

@misc{sun_automatic_2024,
  title = {Automatic {Essay} {Multi}-dimensional {Scoring} with {Fine}-tuning and {Multiple} {Regression}},
  author = {Sun, Kun and Wang, Rong},
  year = {2024},
  publisher = {arXiv},
  doidoi = {10.48550/arXiv.2406.01198},
  urlurl = {http://arxiv.org/abs/2406.01198},
}

@misc{qiu_stella_2025,
  title = {{SteLLA}: {A} {Structured} {Grading} {System} {Using} {LLMs} with {RAG}},
  author = {Qiu, Hefei and White, Brian and Ding, Ashley and Costa, Reinaldo and Hachem, Ali and Ding, Wei and Chen, Ping},
  year = {2025},
  publisher = {arXiv},
  doidoi = {10.48550/arXiv.2501.09092},
  urlurl = {http://arxiv.org/abs/2501.09092},
}

@article{arslan_survey_2024,
  title = {A {Survey} on {RAG} with {LLMs}},
  author = {Arslan, Muhammad and Ghanem, Hussam and Munawar, Saba and Cruz, Christophe},
  journal = {Procedia Computer Science},
  volume = {246},
  pages = {3781--3790},
  year = {2024},
  publisher = {Elsevier BV},
  doidoi = {10.1016/j.procs.2024.09.178},
  urlurl = {https://linkinghub.elsevier.com/retrieve/pii/S1877050924021860},
  issn = {1877-0509},
}

@misc{reimers_sentence-bert_2019,
  title = {Sentence-{BERT}: {Sentence} {Embeddings} using {Siamese} {BERT}-{Networks}},
  author = {Reimers, Nils and Gurevych, Iryna},
  year = {2019},
  publisher = {arXiv},
  doidoi = {10.48550/arXiv.1908.10084},
  urlurl = {http://arxiv.org/abs/1908.10084},
}

@misc{do_autoregressive_2024,
  title = {Autoregressive {Score} {Generation} for {Multi}-trait {Essay} {Scoring}},
  author = {Do, Heejin and Kim, Yunsu and Lee, Gary Geunbae},
  year = {2024},
  publisher = {arXiv},
  doidoi = {10.48550/arXiv.2403.08332},
  urlurl = {http://arxiv.org/abs/2403.08332},
}

@inproceedings{cummins_constrained_2016,
  title = {Constrained {Multi}-{Task} {Learning} for {Automated} {Essay} {Scoring}},
  author = {Cummins, Ronan and Zhang, Meng and Briscoe, Ted},
  booktitle = {Proceedings of the 54th {Annual} {Meeting} of the {Association} for {Computational} {Linguistics} ({Volume} 1: {Long} {Papers})},
  pages = {789--799},
  year = {2016},
  publisher = {Association for Computational Linguistics},
  editor = {Erk, Katrin and Smith, Noah A.},
  doidoi = {10.18653/v1/P16-1075},
  urlurl = {https://aclanthology.org/P16-1075/},
  address = {Berlin, Germany},
}

@misc{rodriguez_language_2019,
  title = {Language models and {Automated} {Essay} {Scoring}},
  author = {Rodriguez, Pedro Uria and Jafari, Amir and Ormerod, Christopher M.},
  year = {2019},
  publisher = {arXiv},
  doidoi = {10.48550/arXiv.1909.09482},
  urlurl = {http://arxiv.org/abs/1909.09482},
}

@article{elmassry_systematic_2025,
  title = {A {Systematic} {Review} of {Pretrained} {Models} in {Automated} {Essay} {Scoring}},
  author = {Elmassry, Ahmed M. and Zaki, Nazar and Alsheikh, Negmeldin and Mediani, Mohammed},
  journal = {IEEE Access},
  volume = {13},
  pages = {121902--121917},
  year = {2025},
  doidoi = {10.1109/ACCESS.2025.3584784},
  urlurl = {https://ieeexplore.ieee.org/abstract/document/11062635},
  issn = {2169-3536},
}

@misc{patil_auto_nodate,
  title = {Auto {Grader} for {Short} {Answer} {Questions}},
  author = {Patil, Pranjal and Agrawal, Ashwin},
  year = {2018},
  howpublished = {Stanford {CS224N} course project report},
}

@misc{gao_simcse_2022,
  title = {{SimCSE}: {Simple} {Contrastive} {Learning} of {Sentence} {Embeddings}},
  author = {Gao, Tianyu and Yao, Xingcheng and Chen, Danqi},
  year = {2022},
  publisher = {arXiv},
  doidoi = {10.48550/arXiv.2104.08821},
  urlurl = {http://arxiv.org/abs/2104.08821},
}

@misc{khattab_colbert_2020,
  title = {{ColBERT}: {Efficient} and {Effective} {Passage} {Search} via {Contextualized} {Late} {Interaction} over {BERT}},
  author = {Khattab, Omar and Zaharia, Matei},
  year = {2020},
  publisher = {arXiv},
  doidoi = {10.48550/arXiv.2004.12832},
  urlurl = {http://arxiv.org/abs/2004.12832},
}

@article{li_enhanced_2022,
  title = {Enhanced hybrid neural network for automated essay scoring},
  author = {Li, Xia and Yang, Huali and Hu, Shengze and Geng, Jing and Lin, Keke and Li, Yuhai},
  journal = {Expert Systems},
  volume = {39},
  number = {10},
  pages = {1--22},
  year = {2022},
  publisher = {Wiley-Blackwell},
  doidoi = {10.1111/exsy.13068},
  urlurl = {https://login.mit.idm.oclc.org/login?url=https://search.ebscohost.com/login.aspx?direct=true&db=buh&AN=160065652&login.asp&site=ehost-live&scope=site},
  issn = {02664720},
}

@article{janda_syntactic_2019,
  title = {Syntactic, {Semantic} and {Sentiment} {Analysis}: {The} {Joint} {Effect} on {Automated} {Essay} {Evaluation}},
  author = {Janda, Harneet Kaur and Pawar, Atish and Du, Shan and Mago, Vijay},
  journal = {IEEE Access},
  volume = {7},
  pages = {108486--108503},
  year = {2019},
  doidoi = {10.1109/ACCESS.2019.2933354},
  urlurl = {https://ieeexplore.ieee.org/document/8788526/},
  issn = {2169-3536},
}

@inproceedings{vaswani_attention_nodate,
  title = {Attention is {All} you {Need}},
  author = {Vaswani, Ashish and Shazeer, Noam and Parmar, Niki and Uszkoreit, Jakob and Jones, Llion and Gomez, Aidan N. and Kaiser, {\L}ukasz and Polosukhin, Illia},
  booktitle = {Advances in {Neural} {Information} {Processing} {Systems} 30 ({NeurIPS} 2017)},
  pages = {5998--6008},
  year = {2017},
}

@misc{dai_semi-supervised_2015,
  title = {Semi-supervised {Sequence} {Learning}},
  author = {Dai, Andrew M. and Le, Quoc V.},
  year = {2015},
  publisher = {arXiv},
  doidoi = {10.48550/arXiv.1511.01432},
  urlurl = {http://arxiv.org/abs/1511.01432},
}

@techreport{radford_language_nodate,
  title = {Language {Models} are {Unsupervised} {Multitask} {Learners}},
  author = {Radford, Alec and Wu, Jeffrey and Child, Rewon and Luan, David and Amodei, Dario and Sutskever, Ilya},
  institution = {OpenAI},
  year = {2019},
}

@misc{latif_fine-tuning_2023,
  title = {Fine-tuning {ChatGPT} for {Automatic} {Scoring}},
  author = {Latif, Ehsan and Zhai, Xiaoming},
  year = {2023},
  publisher = {arXiv},
  doidoi = {10.48550/arXiv.2310.10072},
  urlurl = {http://arxiv.org/abs/2310.10072},
}

@misc{stahl_exploring_2024,
  title = {Exploring {LLM} {Prompting} {Strategies} for {Joint} {Essay} {Scoring} and {Feedback} {Generation}},
  author = {Stahl, Maja and Biermann, Leon and Nehring, Andreas and Wachsmuth, Henning},
  year = {2024},
  publisher = {arXiv},
  doidoi = {10.48550/arXiv.2404.15845},
  urlurl = {http://arxiv.org/abs/2404.15845},
}

@article{hussein_trait-based_2020,
  title = {A {Trait}-based {Deep} {Learning} {Automated} {Essay} {Scoring} {System} with {Adaptive} {Feedback}},
  author = {Hussein, Mohamed A. and Hassan, Hesham A. and Nassef, Mohammad},
  journal = {International Journal of Advanced Computer Science and Applications},
  volume = {11},
  number = {5},
  year = {2020},
  doidoi = {10.14569/IJACSA.2020.0110538},
  urlurl = {http://thesai.org/Publications/ViewPaper?Volume=11&Issue=5&Code=IJACSA&SerialNo=38},
  issn = {21565570, 2158107X},
}

@misc{hou_improve_2025,
  title = {Improve {LLM}-based {Automatic} {Essay} {Scoring} with {Linguistic} {Features}},
  author = {Hou, Zhaoyi Joey and Ciuba, Alejandro and Li, Xiang Lorraine},
  year = {2025},
  publisher = {arXiv},
  doidoi = {10.48550/arXiv.2502.09497},
  urlurl = {http://arxiv.org/abs/2502.09497},
}

@inproceedings{xiao_human-ai_2025,
  title = {Human-{AI} {Collaborative} {Essay} {Scoring}: {A} {Dual}-{Process} {Framework} with {LLMs}},
  author = {Xiao, Changrong and Ma, Wenxing and Song, Qingping and Xu, Sean Xin and Zhang, Kunpeng and Wang, Yufang and Fu, Qi},
  booktitle = {15th {International} {Learning} {Analytics} and {Knowledge} {Conference}},
  pages = {293--305},
  year = {2025},
  publisher = {Association for Computing Machinery},
  doidoi = {10.1145/3706468.3706507},
}

@inproceedings{frederick_eneye_advances_2025,
  title = {{Advances in {Auto}-{Grading} with {Large} {Language} {Models}: {A} {Cross}-{Disciplinary} {Survey}}},
  author = {Frederick Eneye, Tania Amanda Nkoyo and Ijezue, Chukwuebuka Fortunate and Imam Amjad, Ahmad and Amjad, Maaz and Butt, Sabur and Casta{\~n}eda-Garza, Gerardo},
  booktitle = {{Proceedings of the 20th {Workshop} on {Innovative} {Use} of {NLP} for {Building} {Educational} {Applications} ({BEA} 2025)}},
  pages = {477--498},
  year = {2025},
  publisher = {Association for Computational Linguistics},
  doidoi = {10.18653/v1/2025.bea-1.35},
  urlurl = {https://aclanthology.org/2025.bea-1.35/},
  address = {Vienna, Austria},
  isbn = {979-8-89176-270-1},
}

@inproceedings{dzikovska_semeval_2013,
  title     = {{{SemEval-2013} Task 7: The Joint Student Response Analysis and 8th Recognizing Textual Entailment Challenge}},
  author    = {Dzikovska, Myroslava and Nielsen, Rodney and Brew, Chris and Leacock, Claudia and Giampiccolo, Danilo and Bentivogli, Luisa and Clark, Peter and Dagan, Ido and Dang, Hoa Trang},
  booktitle = {{Volume 2: Proceedings of the Seventh International Workshop on Semantic Evaluation (SemEval 2013)}},
  pages     = {263--274},
  year      = {2013},
  publisher = {Association for Computational Linguistics}
}

@inproceedings{doewes_evaluating_2023,
  title     = {Evaluating Quadratic Weighted Kappa as the Standard Performance Metric for Automated Essay Scoring},
  author    = {Doewes, Afrizal and Kurdhi, Nughthoh Arfawi and Saxena, Akrati},
  booktitle = {{Proceedings of the 16th International Conference on Educational Data Mining (EDM)}},
  pages     = {103--113},
  year      = {2023},
  doi       = {10.5281/zenodo.8115784}
}

@article{douze_faiss_2025,
  title     = {The {Faiss} library},
  author    = {Douze, Matthijs and Guzhva, Alexandr and Deng, Chengqi and Johnson, Jeff and Szilvasy, Gergely and Mazar{\'e}, Pierre-Emmanuel and Lomeli, Maria and Hosseini, Lucas and J{\'e}gou, Herv{\'e}},
  journal   = {IEEE Transactions on Big Data},
  year      = {2025},
  publisher = {IEEE},
  doi       = {10.1109/TBDATA.2025.3618474}
}
\end{document}